\documentclass[twocolumn,showpacs,pra,superscriptaddress]{revtex4}
\usepackage{graphicx}
\usepackage{amsmath}

\begin{document}

\title{Non-Markovian effect on the geometric phase of a dissipative qubit}
\author{Juan-Juan Chen}
\affiliation{Center for Interdisciplinary Studies, Lanzhou
University, Lanzhou 730000, China} \affiliation{School of Nuclear
Science and Technology, Lanzhou University, Lanzhou 730000, China }
\author{Jun-Hong An}\email{anjhong@lzu.edu.cn}
\affiliation{Center for Interdisciplinary Studies, Lanzhou
University, Lanzhou 730000, China} \affiliation{Department of
Physics, National University of Singapore, 3 Science Drive 2,
Singapore 117543, Singapore}
\author{Qing-Jun Tong}
\affiliation{Center for Interdisciplinary Studies, Lanzhou
University, Lanzhou 730000, China} \affiliation{School of Nuclear
Science and Technology, Lanzhou University, Lanzhou 730000, China }
\author{Hong-Gang Luo}
\affiliation{Center for Interdisciplinary Studies, Lanzhou
University, Lanzhou 730000, China}
\author{C. H. Oh}
\affiliation{Department of Physics, National University of
Singapore, 3 Science Drive 2, Singapore 117543, Singapore}

\begin{abstract}
We study the geometric phase of a two-level atom coupled to an
environment with Lorentzian spectral density. The non-Markovian
effect on the geometric phase is explored analytically and
numerically. In the weak coupling limit the lowest-order correction
to the geometric phase is derived analytically and the general case
is calculated numerically. It is found that the correction to the
geometric phase is significantly large if the spectral width is
small and in this case the non-Markovian dynamics has a significant
impact to the geometric phase. When the spectral width increases,
the correction to the geometric phase becomes negligible, which
shows the robustness of the geometric phase to the environmental
white noises. The result is significant to the quantum information
processing based on the geometric phase.
\end{abstract}

 \maketitle

%%%%%%%%%%%%%%%%%%%%%%%%%%%%%%%%%%%%%%%%%%%%%%%%%%%%%%%%%%%%%%%%%%%%%%%%%%%%%%%%%%%%%%%%%%%
%%%%%%%%%%%%%%%%%%%%%%%%%%%%%%%%%%%%%%%%%%%%%%%%%%%%%%%%%%%%%%%%%%%%%%%%%%%%%%%%%%%%%%%%%%%
\section{Introduction}\label{Intr}
The concept of geometric phase (GP) in quantum system was originally
introduced by Berry \cite{berry84} when he studied the dynamics of a
closed quantum system which undergoes an adiabatic cyclic evolution.
He found that besides the usual dynamical phase, the system also
acquires an additional phase which only depends on the geometry of
the path traversed by the system during its adiabatic evolution.
Since then this important notion has attracted much attention
\cite{shapere89}. It has been generalized in various aspects, e.g.
the GP for non-adiabatic evolution \cite{Aharonov87} and for
noncyclic evolution \cite{Samuel88}. The GP has been observed
experimentally in optical \cite{Chiao86}, NMR \cite{Du07,Du09}, and superconducting electronic circuit experiments
\cite{Leek07,Mottonen08}.

Recently, the renewed interest in the investigation of GP comes from
the application of GP to implement the logic gates in quantum
computation \cite{Jones00}. The purely geometric nature of the phase
makes such computation intrinsically fault-tolerant and robust
against certain types of classical fluctuation noise
\cite{Zanardi99,Chiara03,Zhu05,Lupo09,Leibfried,Filipp09}. However,
any realistic quantum system is inevitably coupled to its
surrounding environment, which would result in the loss of quantum
coherence (i.e. the decoherence) of the quantum system itself and
hence limit the implementation of the geometric quantum computation.
Therefore, the study of the GP in open quantum systems becomes an
important issue. For the GP of mixed states in open systems, Uhlmann
was the first to make the attempt to define the mixed-state GP via
state purification \cite{Uhlmann}. Sj\"{o}qvist {\it et al.}
proposed an alternative definition for the nondegenerate mixed-state
density matrix under unitary evolution based on the interferometry
\cite{Sjoqvist00}. This definition was further generalized to
degenerate mixed state by Singh {\it et al.} \cite{Singh} and to the
nonunitary evolution by Tong {\it et al.} \cite{tong04} using the
kinematic approach. The GP associated with such a nonunitary
evolution is defined as \cite{tong04}
\begin{eqnarray}
\Phi_g &=&\arg \Big\{ \sum_{k}\sqrt{\varepsilon _{k}(0)\varepsilon
_{k}(T
)}\left\langle \psi _{k}(0)\right\vert \psi _{k}(T )\rangle   \notag \\
&&\times e^{-\int_{0}^{T }dt\left\langle \psi _{k}\right\vert \frac{%
\partial }{\partial t}\left\vert \psi _{k}\right\rangle }\Big\},  \label{GPt}
\end{eqnarray}
where $\varepsilon _{k}(t)$ and $\left\vert \psi _{k}\right\rangle
$, respectively, are the eigenvalues and the eigenstates of the
reduced density matrix of the quantum system, and $T$ is the time
after the system completes a cyclic evolution when it is isolated
from the environment. Taking the environment into account, the
system no longer undergoes a cyclic evolution. Here a quasicyclic
process with $T=2\pi/\omega_0$, where $\omega_0$ is the frequency of
the system, is considered in Eq. (\ref{GPt}). Wang {\it et al.}
defined a mixed-state GP in the context of Pantcharatnam formula
\cite{Pan05} via mapping the density matrix to a nonunit vector ray
in complex projective Hilbert space \cite{Wang06}. The mixed-state
GP has been observed in NMR system \cite{Du03,Du072}.

Because the environment induced decoherence would affect the
performance of the quantum computation using GP, the study of
environment effects on the GP is highly desired. Many works along
this line have been performed within Markovian approximation
\cite{ellinas89,marzlin04,carollo,banerjee08,Buric09,Tong09}, which
is valid only when the interaction between the system and the
environment is very weak and the environmental correlation time is
very small. However, in many quantum information experiments, these
conditions are not completely satisfied. For example, in cavity QED
experiment the imperfection of the cavity mirrors makes the cavity
field having a Lorentzian spectrum expansion, which acts as an
environment, would exert a strong non-Markovian effect on the atom
in it \cite{Bellomo07}. There are also some works on the
non-Markovian effect on the GP in dephasing environments
\cite{yi05,yi06,Lombardo06,dajka08,villar09}, where there is no
energy/information exchange between the system and its environment.

In this work, we extend the study of the GP of open two-level system
to the situation where there has an energy/information exchange
between the system and its environment. The environment is at
zero-temperature and has a Lorentzian spectral density, which
corresponds to the radiation field as an environment being confined
in a leaky cavity. We mainly concentrate on how the non-Markvoain
effect affects the GP in different parameter regimes of the spectral
density, and how the GP manifests its robustness against the
decoherence.

This paper is organized as follows. In Sec. \ref{Mode}, we introduce
the model of a qubit interacting with a vacuum environment and
analyze its decoherence behavior. In Sec. \ref{geometric} we
evaluate the GP of the qubit via performing analytical and numerical
calculations. The correction effect exerted by the environment on
the GP in different parameter regimes of the Lorentzian spectral
density is analyzed. Finally, a brief discussion and summary are
given in Sec. \ref{Diss}.

\section{The model}\label{Mode}
Let us consider a system consisted of a two-level atom (qubit)
coupled to a radiation field at zero temperature as an
environment. The Hamiltonian of the system is \cite%
{Scully97}
\begin{equation}
H=\omega _{0}\sigma _{+}\sigma _{-}+\sum_{k}\omega _{k}a_{k}^{\dag
}a_{k}+\sum_{k}(g_{k}\sigma _{+}a_{k}+h.c.),  \label{t1}
\end{equation}
where $\sigma _{\pm }$ and $\omega _{0}$ are the inversion operators and
transition frequency of the qubit, $a_{k}^{\dag }$ and $a_{k}$ are the
creation and annihilation operators of the $k$-th mode with frequency $%
\omega _{k}$ of the radiation field, and $g_{k}$ is the coupling
strength between the qubit and reservoir. Throughout this paper we
assume $\hbar =1$. The system models the decoherence process of the
atom via the amplitude decaying under the Born-Markovian
approximation, which results in the spontaneous emission of the
two-level atom \cite{Scully97} in quantum optics. The model is
exactly solvable. The decoherence dynamics of the qubit is governed
by the master equation \cite{Breuer02}
\begin{eqnarray}
\dot{\rho}(t) &=&-i\Delta (t)[\sigma _{+}\sigma _{-},\rho
(t)]+\Gamma
(t)[2\sigma _{-}\rho (t)\sigma _{+}  \notag \\
&&-\sigma _{+}\sigma _{-}\rho (t)-\rho (t)\sigma _{+}\sigma _{-}],
\label{mstt}
\end{eqnarray}%
where the time-dependent parameters are given by
\begin{equation}
\Delta (t)=-\text{Im}[\frac{\dot{c}(t)}{c(t)}],~\Gamma (t)=-\text{Re}[\frac{%
\dot{c}(t)}{c(t)}].  \label{coe}
\end{equation}%
It is shown that $c(t)$ satisfies
\begin{equation}
\dot{c}(t)+i\omega _{0}c(t)+\int_{0}^{t}f(t-\tau )c(\tau )d\tau =0,
\label{c0}
\end{equation}
where $f(t-\tau )=\int J(\omega )e^{-i\omega (t-\tau )}d\omega $ is
the environmental correlation function with the spectral density
defined as $J(\omega )=\sum_{k}|g_{k}|^{2}\delta (\omega
-\omega _{k})$ and the initial condition $c(0)=1$. The time-dependent parameters $\Delta (t)$ and $
\Gamma (t)$ play the roles of Lamb shifted frequency and decay rate
of the qubit, respectively. The integro-differential equation
(\ref{c0}) contains the memory effect of the reservoir registered in
the time-nonlocal kernel function and thus the dynamics of qubit
displays non-Markovian effect. If the time-nonlocal kernel function
is replaced by a time-local one, then Eq. (\ref{mstt}) recovers the
conventional master equation under Markovian approximation.

It is obvious that the memory effect registered in the kernel
function $ f(t-\tau)$ is essentially determined by the spectral
density $J(\omega)$. In this work we explicitly consider that the spectral density has a Lorentzian form \cite{Maniscalco08}
\begin{equation}
J(\omega )=\frac{1}{\pi }\frac{W^{2}\lambda }{\left( \omega _{0}-\omega
\right) ^{2}+\lambda ^{2}},  \label{sdensity}
\end{equation}
where $W$ is the coupling constant between the qubit and the
environment, and $\lambda $ defines the spectral width of the
coupling at the resonance point $\omega _{0}$. The Lorentzian
spectral density describes that the vacuum radiation field as the
environment is confined in a leaky cavity. Due to the leakage of the
cavity field induced by the imperfection of the cavity mirrors, the
spectrum of the cavity field displays a broadening at the resonance
point regarding the atomic transition frequency $\omega_0$. In this
case one can verify that the correlation function decays
exponentially $ f\left( t-\tau \right) =W^{2}e^{-\lambda (t-\tau )}$
\cite{Maniscalco08}, which means that the parameter $\lambda $
characterizes the correlation time of the environment as $\tau
_{c}=\lambda ^{-1}$. If $\tau_c$ is comparable with the typical time
scale of the system, i.e. $\tau_0=1/\omega_0$, then the memory
effect of the environment should not be neglected and the
decoherence dynamics in this situation is non-Markovian. While
$\tau_c\ll \tau_0$, the memory effect of the environment is
negligible and the decoherence dynamics is Markovian. In the ideal
cavity limit $\lambda \rightarrow 0$, we have $\lim_{\lambda
\rightarrow 0}J(\omega )=W^{2}\delta \left( \omega -\omega
_{0}\right)$, which corresponds to a constant kernel $f\left( t-\tau
\right) =W^{2}$. Then the system reduces to the Jaynes-Cummings
model with a vacuum Rabi frequency $g=W$.

Going back to the general case, one can obtain the analytical form
of $c(t)$ by substituting the exponentially decaying $f(t-\tau)$
into Eq. (\ref{c0}) as
\begin{equation}\label{cf}
c(t)=e^{-\frac{(\lambda+i\omega_0) t}{2}}\left[ \cosh (\frac{\Omega t
}{2})+\frac{\lambda }{\Omega }\sinh (\frac{\Omega t}{2})\right],
\end{equation}where $\Omega =\sqrt{\lambda ^{2}-4W^{2}}$. So the parameters in the master
equation can be calculated readily
\begin{equation}\label{gf}
\Gamma(t)=\frac{2W^2}{\lambda+\Omega\coth(\Omega
t/2)},~~\Delta(t)=\omega_0.
\end{equation}
One can see from Eqs. (\ref{gf}) when $\lambda $ is much larger than
other frequency scale, the decay rate tends to a constant value
$\Gamma_0 \equiv2W^{2}/\lambda $, which just characterizes the
decoherence behavior of the qubit under the Markovian dynamics.

\section{Geometric phase corrected by the environment}\label{geometric}
\subsection{Analytical analysis}
In the following we compute explicitly the GP of the qubit. We
assume that the initial state of the qubit is chosen as
\begin{equation}
\left\vert \psi (0)\right\rangle =\cos \frac{\theta _{0}}{2}\left\vert
+\right\rangle+\sin \frac{\theta _{0}}{2}\left\vert -\right\rangle ,
\label{istate}
\end{equation}
where $|+\rangle$ and $|-\rangle$ are the excited and ground states
of the qubit, respectively. This state corresponds to a vector in
Bloch sphere with polar angle $\theta_0$. The time-dependent reduced
density matrix of the qubit under the initial condition
(\ref{istate}) can be obtained straightforwardly from the master
equation (\ref{mstt})
\begin{equation}
\rho (t)=\left(
\begin{array}{cc}
\cos ^{2}\frac{\theta _{0}}{2}|c(t)|^{2} & \frac{\sin \theta _{0}}{2}c(t) \\
\frac{\sin \theta _{0}}{2}c^*(t) & 1-\cos ^{2}\frac{\theta _{0}}{2}|c(t)|^{2}%
\end{array}
\right).  \label{rdm}
\end{equation}
It is noted that besides the off-diagonal elements, the diagonal
elements of the reduced density matrix $\rho(t)$ also change with
time in our model. It is just this time-dependence of the diagonal
elements of $\rho(t)$ characterizing the energy exchange between the
qubit and its environment that makes our system shows dramatic
difference to the dephasing model
\cite{yi05,yi06,Lombardo06,dajka08,villar09}. To calculate the GP of
the qubit, we must firstly get the eigensolution of the reduced
density matrix (\ref{rdm}). The eigenvalues of the above reduced
density are readily calculated,
\begin{equation}
\varepsilon _{\pm }(t) =\frac{1}{2}\left[ 1\pm \sqrt{|c(t)|^{2}\sin
^{2}\theta _{0}+\left( 2|c(t)|^{2}\cos ^{2}\frac{\theta _{0}}{2}-1\right)
^{2}}\right] ,  \label{eigenv}
\end{equation}
It is obvious that the eigenvalue $\varepsilon_-(0)=0$, which, from Eq. (\ref%
{GPt}), means that the component of the eigenstate corresponding to
the eigenvalue $\varepsilon_-$ gives no contribution to the GP. Thus
we only need to consider the eigenstate corresponding to the
eigenvalue $\varepsilon_+$
\begin{equation}
\left\vert \varepsilon _{+}(t)\right\rangle =e^{-i\omega _{0}t}\cos \Theta
\left\vert +\right\rangle +\sin \Theta \left\vert -\right\rangle ,
\label{eigens}
\end{equation}%
where%
\begin{equation}
\cos \Theta =\frac{2\left( |c(t)|^{2}\cos ^{2}\frac{\theta _{0}}{2}%
-\varepsilon _-\right) }{\sqrt{|c(t)|^{2}\sin ^{2}\theta _{0}+4\left(
|c(t)|^{2}\cos ^{2}\frac{\theta _{0}}{2}-\varepsilon _-\right)^2 }}.
\label{angle}
\end{equation}

\begin{figure}[h]
\begin{center}
\includegraphics[width = 0.45\columnwidth]{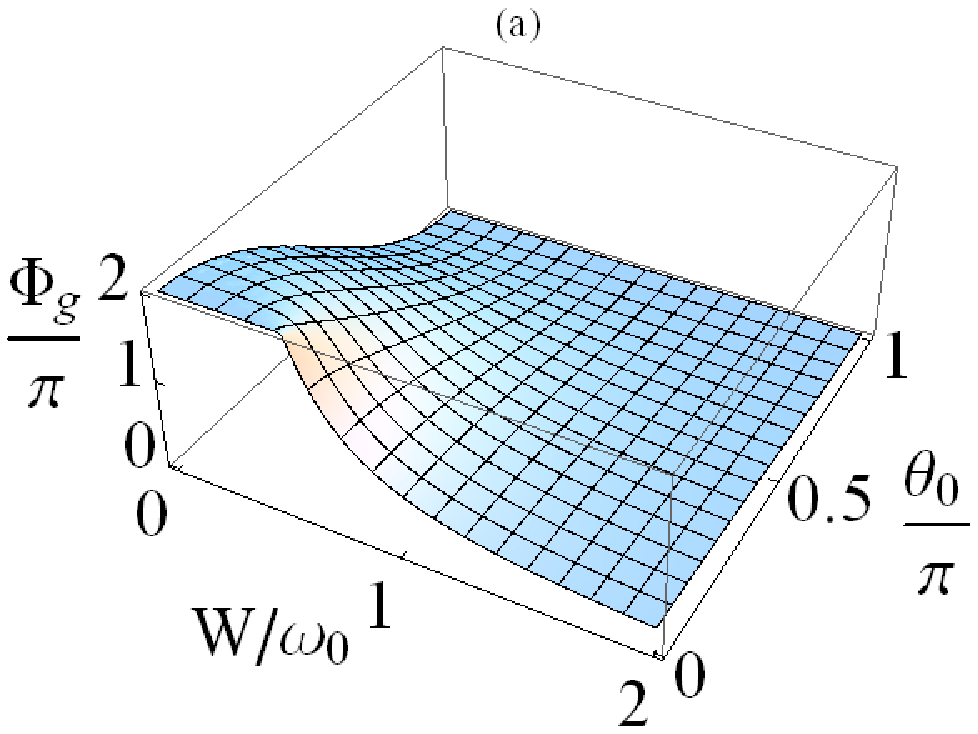}
\includegraphics[width = 0.45\columnwidth]{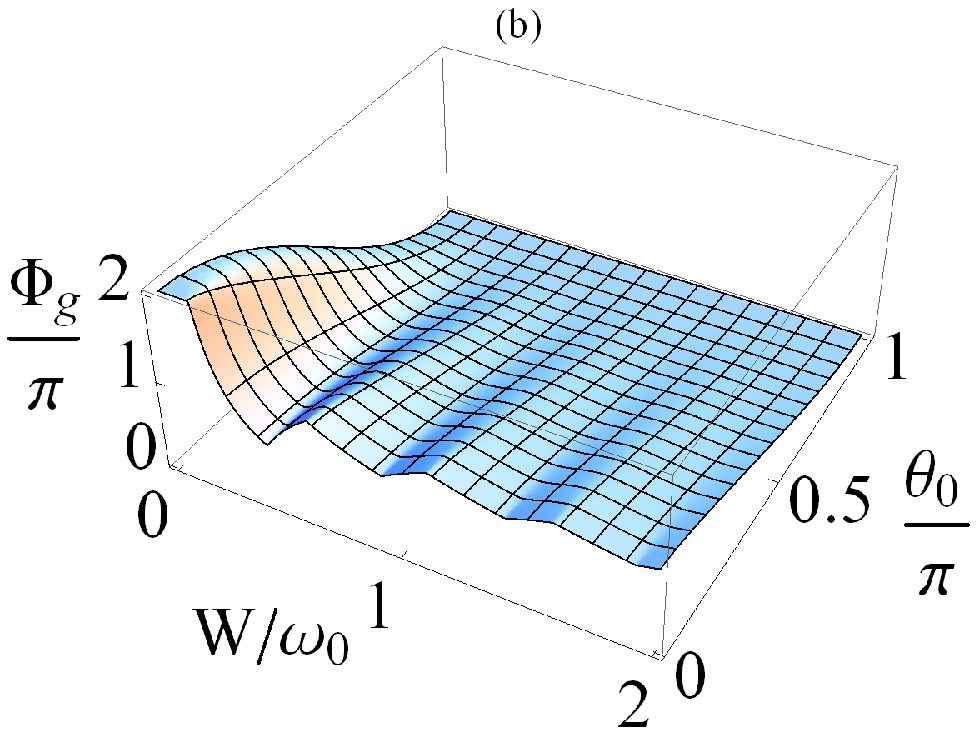}
\end{center}
\caption{(Color online) Numerical result of the GP as a function of the coupling
constant and the initial angle. (a) Markovian dynamics with $\lambda=5.0\omega_0 $ and (b) non-Markovian dynamics with
$\lambda=0.05\omega_0 $. } \label{sle}
\end{figure}

Below we calculate the GP. Eq. (\ref{gf}) shows that the
frequency shift of the qubit is zero for the Lorentzian spectral
density \cite{Breuer02}. So the period of the environment disturbed
atom is the same as the one for a bare atom. Then the GP of the qubit
acquired after a period $T=2\pi /\omega _{0}$ can be calculated as
\begin{equation}
\Phi _{g}=\int_{0}^{T}\omega _{0}\cos ^{2}\Theta dt. \label{GP2}
\end{equation}
Here we point out that the expression of the GP (\ref{GP2}) has been
obtained by using the kinematic approach \cite{tong04}. However, it
can be shown that the expression is exactly the same as the one
obtained by the definition of Pantcharatnam's phase of a nonunit
vector ray \cite{Wang06}, which exhibits some essential features of
the GP in such systems.

From Eq. (\ref{GP2}), one can notes that the effect of the
environment enters into the GP solely via the time-dependent factor
$|c(t)|^2$ of the excited-state population. When the environment is
absent, then $|c(t)|^2=1$ and the GP reduces to $\Phi
_{g}^{(0)}\equiv \pi \left( 1+\cos \theta _{0}\right)$, which is
just the GP acquired by a two-level atom in the unitary dynamics.
Depending on the decoherence dynamics being Markovian or
non-Markovian, the short-time dynamics of $|c(t)|^2$ shows
remarkably different behaviors. In the Markovian dynamics,
$|c(t)|^2$ decays monotonically and finally approaches zero. In this
case, the larger the decay rate $\Gamma_0$ is, the larger the
correction of the GP should be. In the non-Markovian dynamics,
contributed from the memory effect of the environment $|c(t)|^2$
shows transient oscillation with time, which naturally induces a
correction of the GP different to the one in Markovian dynamics, as
shown in the following.

Up to the first order of coupling strength $W^{2}$, i.e. the weak coupling limit, we have%
\begin{equation}
\Phi _{g}^{(1)} =\Phi_g^{(0)} -\left( \frac{W}{%
\omega _{0}}\right) ^{2}\sin ^{2}\theta _{0}\left( 1+\frac{\cos \theta _{0}}{
2}\right)z\left( \frac{\lambda }{\omega _{0}}\right)  , \label{eGP}
\end{equation}
where $z\left( x\right) \equiv x ^{-3}\left[ 1-e^{-2\pi x}-%
2\pi x\left( 1-\pi x \right) \right] $. Besides the leading term
$\Phi_g^{(0)}$ corresponding to the well-known GP acquired under the
unitary dynamics, the second term, which is quadratic in
$W/\omega_0$ with a $\lambda$-dependent coefficient
$z(\lambda/\omega_0)$, is the lowest-order correction to the GP
induced by non-unitary dynamics due to the interaction with the
environment. It is easy to check that $z(\lambda/\omega_0)$ is a
monotonically decreasing function with the increase of $\lambda$, so
we can expect that the GP shows a larger deviation to $\Phi_g^{(0)}$
for a small $\lambda$ than a large one. In particular, in the ideal
cavity limit $\lambda \rightarrow 0$, the function
$z(\lambda/\omega_0)$ arrives at its maximum $4\pi^3/3$, where the
GP has a largest correction in this weak coupling (or small $W$)
regime. On the other hand, when $\lambda\gg \omega_0,~W$, one can
verify $z(\lambda/\omega_0)\rightarrow 2\pi^2\omega_0/\lambda$.
Consequently, the GP in this Markovian limit is
\begin{equation}
\Phi _{g}^{(1)} =\Phi_g^{(0)} -\frac{\pi^2\Gamma}{\omega_0}\sin ^{2}\theta _{0}\left( 1+\frac{\cos \theta _{0}}{%
2}\right) , \label{eGPm}
\end{equation}
which shows very similar form to the result of \cite{marzlin04}
obtained from the Born-Markovian master equation.

\begin{figure}[h]
\begin{center}
\includegraphics[width = \columnwidth]{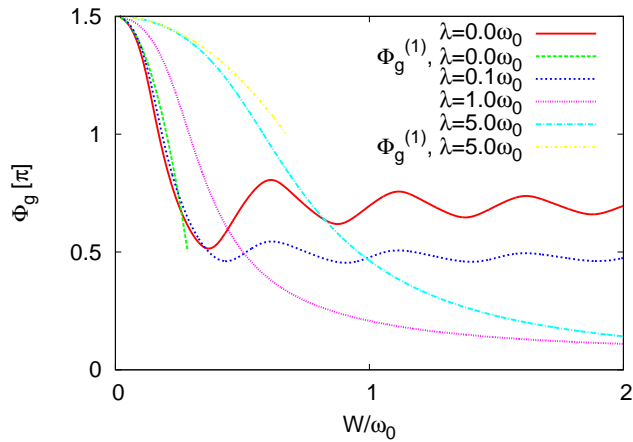}
\end{center}
\caption{(Color online) The exact GP and $\Phi_g^{(1)}$ as a
function of the coupling constant under different spectral width $
\lambda $. The initial polar angle is taken as $\theta_0=\pi/3 $.  }
\label{td}\end{figure}
\begin{figure}[h]
\begin{center}
\includegraphics[width = \columnwidth]{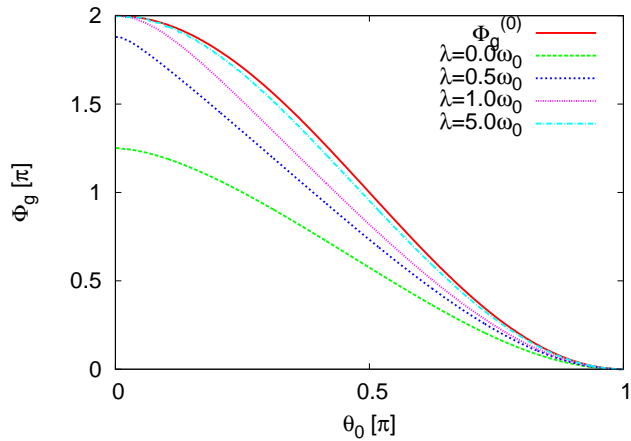}
\end{center}
\caption{(Color online) The exact GP and $\Phi_g^{(0)}$  as a
function of the initial polar angle under different spectral width $
\lambda$.  The coupling constant is $W=0.2\omega_0$. } \label{ws}
\end{figure}

\subsection{Numerical results}
In the following we give the exact GP via numerically evaluating the
integration in Eq. (\ref{GP2}). In Fig. \ref{sle} we plot the exact
GP as a function of the coupling constant and the initial polar
angle. When $\lambda=5.0\omega_0$, the correlation time of the
environment $\tau_c=0.2/\omega_0$ is much less than the typical time
scale of the system $\tau_0$. In this situation, the memory effect
of the environment is negligible and the dynamics of the system is
Markovian. Consequently, the energy and the information flow
single-directionally from the system to the environment, which
results in the dissipation to the system. The dissipation time scale
in this Markovian dynamics is characterized by
$\tau_D=1/\Gamma_0=\lambda/2W^2$. We can see in Fig. \ref{sle}(a)
that the GP in this situation decreases monotonically with the
increase of the coupling constant. This is understandable that a
larger coupling constant would induce a stronger dissipation and a
short decoherence time scale $\tau_D$ to the qubit system. With the
increase of $W$, $\tau_D$ approaches $T$, which means that the
dissipation becomes more and more notable within the time scale $T$.
On the other hand, when $W$ is small the GP shows small deviation to
the unitary one $\Phi_g^{(0)}$. In this situation, $\tau_D$ is
larger than $T$, making the dissipation negligible within the time
scale $T$ during which the GP is accumulated. This interesting
phenomenon just reflects the resilient ability of the GP to the
environment's disturbance, especially in weak coupling limit. A
similar phenomenon is also observed in the dephasing model
\cite{villar09}. When $\lambda=0.05\omega_0$, $\tau_c\gg \tau_0$ and
the non-Markovian effect induced by the memory effect of the
environment to the system would show its distinct impact on the the
dynamics of the system. Besides the dissipation, the environment
also exerts a dynamical backaction on the system \cite{An09}. The
backaction effect is reflected in that the energy and the
information flow back and forth between the system and the
environment. In this situation we can see in Fig. \ref{sle}(b) that
the GP oscillates with the increase of $W$ and then tends to a
definite value, which is qualitatively different to the above
Markovian situation. The oscillation is just the manifestation of
the dynamical backaction. It is very clear that the non-zero
character of the GP in the case of large $W$ is due to the
counteracted competition between the backaction and dissipation
exerted by the non-Markovian environment partially, which weakens
the decoherence of the qubit system.

A more clear comparison of the GP with different $\lambda$ as a
function of $W$ when $\theta_0=\pi/3$ is shown in Fig. \ref{td}. As
comparisons, the perturbed results for $\lambda = 0$ and
$5.0\omega_0$ are also presented in this figure and they agree well
with the corresponding exact one only in the weak coupling limit. We
can see obviously that in the weak coupling regime the environment
with a large $\lambda$ induces a smaller correction to the GP than
the one with a small $\lambda$. That is, the non-Markovian effect
has a strong correction to the GP in this weak coupling regime. This
is consistent with the result obtained in Ref. \cite{Yi08}, where a
phenomenological analysis of the non-Markovian effect on the GP is
performed. While in the strong coupling regime the situation is
opposite, because the system with a large-$\lambda$ shows stronger
decoherence than the one with a small $\lambda$ where the
non-Markovian effect has a strong correction to the GP.

Even till today it is still difficult to access experimentally the
strong coupling regime in the cavity QED platform. For example, in a
recent microtoroidal resonator experiment \cite{Dayan}, the achieved
coupling strength between the atom and the cavity field is 90MHz,
while the damping rate of the cavity field is about 180MHz. Such a
large damping rate would induce a noticeable spectrum expansion of
the cavity mode. It is of course interesting to examine the GP in
this bad-cavity and weak coupling regime. In Fig. \ref{ws} we plot
the GP as a function of the initial polar angle in the weak coupling
regime for different spectrum width $\lambda$. One can see that the
GP approach more and more closely the unitary GP $\Phi_g^{(0)}$ with
the increase of $\lambda$. This accounts for once again that the GP
shows more strong resilient ability to the noise of the Markovian
environment with a large $\lambda$. Since this resilient character
is obtained in weak coupling and bad-cavity condition, we argue that
it is accessible by modern cavity QED experiment \cite{Dayan}.

\section{Summary and Discussion} \label{Diss}
We have studied exactly the GP for a qubit in an amplitude decaying
environment using the kinematic approach. We have evaluated the
non-Markovian effect on the GP. It is demonstrated that the
non-Markovian environment has a more significant correction to the
GP in the weak coupling regime due to the strong short-time
correlation between the qubit and the environment. Interestingly,
our result also indicates the insensitivity of the GP to the
Lorentzian environment in the Markovian regime when the coupling is
weak. It just manifests the resilient ability of the GP to the
environmental noise. This results elucidate that the GP in this
cavity QED system is fault-tolerant not only against the classical
noise induced by the parameter fluctuation
\cite{Zanardi99,Leibfried}, but also against the quantum noise. This
has significant meaning in geometric quantum information processing.

Our model, as a basic model of quantum optics, is particularly
relevant to the cavity QED experiments. In this respect some
remarkable experiments, such as the efficient coupling of the
trapped atoms with cavity field \cite{Kimble05,Rempe07}, have been
performed successfully. Practically, any phase variation is observed
only via some kind of interferometry between the involved state and
certain selected reference state. For example, the GP for mixed
state has been observed via designing a quantum network in NMR
system to realize the interferometry \cite{Du03,Du072}. This
provides a clue to observe GP in cavity QED system. Although the GP
has not been observed in the cavity QED system, a quantum network
using the recent developed microtoroidal resonator \cite{Dayan} has
been proposed \cite{Kimble} based on the input-output process of
photons \cite{An092}. If an effective interferometry could be
realized in this quantum network, then the GP would be expected to
be observable in cavity QED system. Our work on the assessment of
the environmental effect on the GP, especially in non-Markovian
regime, is of great importance in using the GP in cavity QED system
to implement the quantum gates.

\section*{Acknowledgement}
This work is supported by NSF of China under Grant No. 10604025,
Gansu Provincial NSF of China under Grant No. 0803RJZA095, the
Program for NCET, and NUS Research Grant No. R-144-000-189-305.


\begin{thebibliography}{99}
\bibitem{berry84} M. V. Berry, Proc. R. Soc. London, Ser. A \textbf{329}, 45
(1984).

\bibitem{shapere89} A. Shapere and F.
Wilczek, {\it Geometric Phases in Physics}, (World Scientific,
Singapore, 1989).

\bibitem{Aharonov87} Y. Aharonov and J. Anandan, Phys. Rev. Lett.
{\bf 58}, 1593 (1987).

\bibitem{Samuel88} J. Samuel and R. Bhandari, Phys. Rev. Lett. {\bf
60}, 2339 (1988).

\bibitem{Chiao86} A. Tomita and R. Y. Chiao, Phys. Rev. Lett. {\bf 57}, 937
(1986).

\bibitem{Du07} J. Du, J. Zhu, M. Shi, X. Peng, and D. Suter, Phys. Rev. A {\bf 76}, 042121
(2007).
\bibitem{Du09} H. Chen, M. Hu, J. Chen, and J. Du, Phys. Rev. A {\bf 80}, 054101
(2009).

\bibitem{Leek07} P. J. Leek, J. M. Fink, A. Blais, R. Bianchetti, M. G\"{o}ppl,
J. M. Gambetta, D. I. Schuster, L. Frunzio, R. J. Schoelkopf, and A. Wallraff, Science {\bf 318}, 1889 (2007).

\bibitem{Mottonen08} M. M\"{o}tt\"{o}nen, J. J. Vartiainen, and J. P. Pekola,
Phys. Rev. Lett. {\bf 100}, 177201 (2008).

\bibitem{Jones00} J. A. Jones, V. Vedral, A. Ekert, and G.
Castagnoli, Nature (London) {\bf 403}, 869 (2000).

\bibitem{Zanardi99} P. Zanardi and M. Rasetti, Phys. Lett. A {\bf
264}, 94 (1999).

\bibitem{Chiara03} G. De Chiara and G. M. Palma, Phys. Rev. Lett. {\bf 91}, 090404
(2003).

\bibitem{Zhu05} S.-L. Zhu and P. Zanardi, Phys. Rev. A {\bf 72}, 020301(R)
(2005).

\bibitem{Lupo09} C. Lupo and P. Aniello, Phys. Scr. {\bf 79}, 065012
(2009).

\bibitem{Leibfried} D. Leibfried, B. DeMarco, V. Meyer, D. Lucas, M. Barrett, J.
Britton, W. M. Itano, B. Jelenkovi, C. Langer, T. Rosenband, and D.
J. Wineland, Nature (London) {\bf 422}, 412 (2003).

\bibitem{Filipp09} S. Filipp, J. Klepp, Y. Hasegawa, C. Plonka-Spehr, U. Schmidt, P. Geltenbort, and H.
Rauch, Phys. Rev. Lett. {\bf 102}, 030404 (2009).

\bibitem{Uhlmann} A. Uhlmann, Rep. Math. Phys. {\bf 24}, 229 (1986);
Ann. Phys. {\bf 46}, 63 (1989); Lett. Math. Phys. {\bf 21}, 229
(1991).

\bibitem{Sjoqvist00} E. Sj\"{o}qvist, A. K. Pati, A. Ekert, J. S.
Anandan, M. Ericsson, D. K. L. Oi, and V. Vedral, Phys. Rev. Lett.
{\bf 85}, 2845 (2000).

\bibitem{Singh} K. Singh, D. M. Tong, K. Basu, J. L. Chen, and J. F.
Du, Phys. Rev. A {\bf 67}, 032106 (2003).


\bibitem{tong04} D. M. Tong, E. Sj\"{o}qvist, L. C. Kwek, and C. H. Oh,
Phys. Rev. Lett. \textbf{93}, 080405 (2004).

\bibitem{Pan05} S. Pancharatnam, Proc. Indian Acad. Sci., Sect. A
{\bf 44}, 247 (1956).

\bibitem{Wang06} Z. S. Wang, L. C. Lwek, C. H. Lai, and C. H. Oh,
Europhysics Lett. {\bf 74}, 958 (2006).

\bibitem{Du03} J. Du, P. Zou, M. Shi, L. C. Kwek, J. W. Pan, C. H. Oh, A. Ekert, D. K. L. Oi, and M. Ericsson,
Phys. Rev. Lett. \textbf{91}, 100403 (2003).

\bibitem{Du072} J. Du, M. Shi, J. Zhu, V. Vedral, X. Peng, and D.
Suter, arXiv:0710.5804v1 [quant-ph].

\bibitem{ellinas89} D. Ellinas, S. M. Barnett, and M. A. Dupertuis, Phys.
Rev. A \textbf{39}, 3228 (1989).

\bibitem{marzlin04} K.-P. Marzlin, S. Ghose, and B. C. Sanders, Phys. Rev.
Lett. \textbf{93}, 260402 (2004).

\bibitem{carollo} A. Carollo, I. Fuentes-Guridi, M. F. Santos, and V.
Vedral, Phys. Rev. Lett. \textbf{90}, 160402 (2003); \textbf{92},
020402 (2004).

\bibitem{banerjee08}  S. Banerjee and R. Srikanth, Eur. Phys. J. D \textbf{46}%
, 335 (2008).
\bibitem{Buric09} N. Buri\'{c} and M. Radonji\'{c}, Phys. Rev. A
{\bf 80}, 014101 (2009).

\bibitem{Tong09} S. Yin and D. M. Tong, Phys. Rev. A {\bf 79},
044303 (2009).

\bibitem{Bellomo07} B. Bellomo, R. Lo Franco and G. Compagno, Phys. Rev.
Lett. \textbf{99} 160502 (2007).

\bibitem{yi05} X. X. Yi, L. C. Wang, and W. Wang, Phys. Rev. A \textbf{71},
044101 (2005).

\bibitem{yi06} X. X. Yi, D. M. Tong, L. C. Wang, L. C. Kwek, and C. H.
Oh, Phys. Rev. A {\bf 73}, 052103 (2006).

\bibitem{Lombardo06} F. C. Lombardo and P. I. Villar, Phys. Rev. A
{\bf 74}, 042311 (2006).

\bibitem{dajka08} J. Dajka, M. Mierzejewski, and J. {\L}uczka, J. Phys. A: Math.
Theor. \textbf{41}, 012001 (2008).

\bibitem{villar09} P. I. Villar, Phys. Lett. A \textbf{373} (2009).

\bibitem{Scully97} M. O. Scully and M. S. Zubairy, \textit{Quantum Optics}
(Cambridge University Press, Cambridge, 1997).

\bibitem{Breuer02} H.-P. Breuer and F. Petruccione, \textit{The Theory of
Open Quantum Systems} (Oxford University Press, Oxford, 2002).

\bibitem{Maniscalco08} S. Maniscalco, F. Francica, R. L. Zaffino, N. Lo Gullo,
and F. Plastina, Phys. Rev. Lett. {\bf 100}, 090503 (2008).

\bibitem{An09} J.-H. An, Y. Yeo, and C. H. Oh, Ann. Phys. (N.Y.) {\bf 324},
1737 (2009).

\bibitem{Yi08} X. L. Huang and X. X. Yi, Europhysics Lett. {\bf 82}, 50001
(2008).

\bibitem{Dayan} B. Dayan, A. S. Parkins, T. Aoki, E. P. Ostby, K. I.
Vahala, and H. J. Kimble, Science {\bf 319}, 1062 (2008).

\bibitem{Kimble05} K. M. Birnbaum, A. Boca, R. Miller, A. D. Boozer, T.
E. Northup, and H. J. Kimble, Nature (London) {\bf 436}, 87 (2005).

\bibitem{Rempe07} T.Wilk, S. C. Webster, A. Kuhn, and G. Rempe, Science {\bf 317}, 488
(2007).

\bibitem{Kimble} H. J. Kimble, Nature (London) {\bf 453}, 1023 (2008).

\bibitem{An092} J.-H. An, M. Feng, and C. H. Oh, Phys. Rev. A {\bf 79}, 032303
(2009).
\end{thebibliography}
\end{document}